\documentclass[prd,showpacs,11pt]{revtex4}
\usepackage{amsmath}
\usepackage{slashed}
\usepackage{amssymb}
\usepackage{epsfig}
\usepackage{graphicx}
\usepackage{array}
\usepackage{multirow}
\usepackage{color}
\usepackage{txfonts}
\allowdisplaybreaks[4]
\usepackage[colorlinks, citecolor=blue, anchorcolor=red,
menucolor=red, linkcolor=red, filecolor=red, urlcolor=blue, frenchlinks=red]{hyperref}

\begin{document}
\newcommand{\psl}{ p \hspace{-1.8truemm}/ }
\newcommand{\nsl}{ n \hspace{-2.2truemm}/ }
\newcommand{\vsl}{ v \hspace{-2.2truemm}/ }
\renewcommand{\arraystretch}{1.5}
\title{Quasi-two-body $B_{(s)}\to V \pi\pi$ Decays with Resonance $f_0(980)$ in~PQCD Approach }
\author{Lei Yang$^1$}
\author{Zhi-Tian Zou$^1$}
\author{Ying Li$^{1,2}$} \email{liying@ytu.edu.cn}
\author{Xin Liu$^3$}
\author{Cui-Hua Li$^4$}
\affiliation
{$1$ Department of Physics, Yantai University, Yantai 264005, China\\
 $2$ Center for High Energy Physics, Peking University, Beijing 100871, China\\
 $3$ Department of Physics, Jiangsu Normal University, XuZhou 221116, China\\
 $4$ Yantai Engineering and Technology College,Yantai, 264006, China }
\date{\today}
\begin{abstract}
Motivated by the measurements of branching fractions of the quasi-two-body decays $B^0\to K^{*0}(f_0(980)\to )\pi^+\pi^-$ and $B^0\to \rho^0(f_0(980)\to )\pi^+\pi^-$, we study the charmless  $B\to V (f_0(980)\to ) \pi^+\pi^-$ decays in the perturbative QCD approach. Supposing that $f_0(980)$ is a two-quark state and mixture of $n\bar n=(u\bar u+d\bar d)/\sqrt 2$ and $s\bar s$ with the mixing angle $\theta$, we calculate the branching fractions of these decays with the new introduced $s$-wave $\pi^+\pi^-$-pair wave function. When the mixing angle $\theta$ lies in the range $[135^{\circ}, 155^{\circ}]$, the calculated branching fractions of the $B^0\to K^{*0}f_0(980)\to K^{*0}\pi^+\pi^-$ and $B^0\to \rho^0f_0(980)\to \rho^0\pi^+\pi^-$ decays are in agreement with the experimental data. The branching fractions of other decays could be measured in the current LHCb and Belle II experiments.  Considering the isospin symmetry, we also estimate the branching fractions of the quasi-two-body decays $B\to Vf_0(980)\to V\pi^0\pi^0$, which are half of those the corresponding decays $B\to Vf_0(980)\to V\pi^+\pi^-$. Moreover, the direct $CP$ asymmetries of these decays are also calculated, and some of them can be tested in the current experiments.
\end{abstract}
\maketitle
\section{Introduction}
In recent years, non-leptonic three-body decays of $B$ mesons have been paid more attention on both experimental and theoretical sides, as these decays can be used to test the standard model (SM), to extract the CKM angles,  and to search for the sources of the $CP$ violation. From last century, a large number of three-body $B$ decays have been measured by BaBar \cite{Lees:2012kxa}, Belle \cite{Nakahama:2010nj}, CLEO \cite{Eckhart:2002qr} and LHCb \cite{Aaij:2018rol, Aaij:2016qnm, Aaij:2019nmr, Aaij:2017zgz, Aaij:2019jaq, Aaij:2020ypa,Aaij:2020dsq}.  Meanwhile, on the factorization hypothesis, a few of theoretical methods have been proposed to study these decays, such as approaches based on the symmetry principle \cite{He:2014xha}, the QCD factorization approach\cite{Krankl:2015fha, Virto:2016fbw,Cheng:2007si, Cheng:2016shb, Li:2014oca}, the perturbative QCD approach (PQCD) \cite{Wang:2016rlo, Li:2016tpn, Zou:2020atb, Zou:2020fax, Zou:2021lex, Zou:2020dpg}, and other theoretical methods \cite{Wang:2015ula}.

Unlike the two-body decays where the kinematics is fixed, three-body decay amplitudes depend on two kinematic variables. For a decay  $B(p_B) \to M_1(p_1)M_2(p_2)M_3(p_3)$, it is general to define the variables as two invariant masses of two pairs of final state particles, for instance,  $s_{12}$ and $s_{13}$ with the definition $s_{ij}=2(p_i\cdot p_j)/m_B^2$. All physical kinematics configurations could define a two-dimensional region in the $s_{12}-s_{13}$ plane, and the density plot of the differential decay rate $d\Gamma/ ds_{12} ds_{13}$ in this region is called a Dalitz plot. Specially, when the final states are the light mesons such as the $\pi$ and $K$ mesons, the corresponding configuration reduces to a triangle region. In general, the Dalitz plot has three typical regions according to the characteristic kinematics. The central region so-called ``Mercedes Star'' configuration corresponds to the case where all the invariant masses are roughly the same and of order of $m_B$. In this region all three light mesons have a large energy in the $B$ meson rest frame and fly apart at about $120^{\circ}$ angles. The corners regions correspond to the cases where one final state is soft and the others fly back-to-back with large energy about $m_B/2$. At the edges of the Dalitz plot one invariant mass is small and the other two are large, which implies that two particles move collinearly and the third bachelor particle recoils back. The interactions between two collinear mesons leads eventually to the resonances. Compared with two other regions, the physics picture at the edges of the Dalitz plot is very similar to a two-body decay by viewing the two-meson pair as a whole, and we thus call it quasi-two-body decay. In the past twenty years, PQCD approach based on the $k_T$ factorization has been used to study the $B$ meson two-body decays successfully, therefore it can be generalized for studying the quasi-two-body decays.

In the past few years, a large number of charmless quasi-two-body  $B/B_s\to V (f_0(980)\to)\pi^+\pi^-$ decays have been measured in the experiments \cite{Zyla:2020zbs}, and some branching fractions or upper limits of them are summarized as follows
\begin{eqnarray}
B^0\to K^*(892)^0f_0(980),f_0(980)\to \pi^+\pi^-&=&(3.9_{-1.8}^{+2.1})\times10^{-6},\nonumber\\
B^0\to \omega f_0(980),f_0(980)\to \pi^+\pi^-&<&1.5\times10^{-6},\nonumber\\
B^0\to \phi f_0(980), f_0(980)\to \pi^+\pi^-&<&3.8\times 10^{-7},\nonumber\\
B^0\to \rho^0f_0(980),f_0(980)\to \pi^+\pi^-&=&(7.8\pm2.5)\times10^{-7},\nonumber\\
B^+\to \rho^+f_0(980),f_0(980)\to \pi^+\pi^-&<&2.0\times 10^{-6},\nonumber\\
B_s\to \phi f_0(980),f_0(980)\to \pi^+\pi^-&=&(1.12\pm0.21)\times 10^{-6}.
\end{eqnarray}
Except the decay $B_s\to \phi f_0,f_0\to \pi^+\pi^-$, other decays have not been studied theoretically in the literatures. Motivated by this, we shall study above decays in PQCD approach, so as to further check the reliability of PQCD in multi-body decays and present more predictions.For the sake of convenience $f_0(980)$ is abbreviated to $f_0$ in the following context unless special statement.

\section{Framework}
In the framework of PQCD, the decay amplitude $\mathcal{A}$ of $B \to V (f_0(980)\to)\pi^+\pi^-$ decay can be decomposed as the convolution
\begin{eqnarray}\label{convolution}
\mathcal{A}=C(t)\otimes \mathcal{H}(x_i,b_i,t)\otimes \Phi_B(x_1,b_1)\otimes\Phi_{V}(x_2,b_2)
\otimes\Phi_{\pi\pi}(x_3,b_3)\otimes e^{-S(t)},
\end{eqnarray}
where $x_i$ are the momentum fractions of the light quarks, $b_i$ are the conjugate variables of the quarks' transverse momenta $k_{iT}$. $\Phi_B$ and $\Phi_{V}$ are the wave functions of the $B$ mesons and vector mesons, while the $\Phi_{\pi\pi}$ is the $S$-wave $\pi\pi$-pair wave function. These wave functions are non-perturbative and universal. The exponential term is the so-called Sudakov form factor caused by the additional scale introduced by the intrinsic transverse momenta $k_T$, which suppresses the soft dynamics effectively \cite{Li:2001ay,Lu:2000hj}. $\mathcal{H}(x_i,b_i,t)$ is the hard kernel, which can be calculated perturbatively. The parameter $t$ is the largest scale in the hard kernel, which ensures the higher order corrections as small as possible.

In PQCD, the most important inputs are the initial and final mesons' wave functions. For the $B$ meson and the light vector mesons, their wave functions have been studied extensively and the inner parameters have been fixed by the well measured two-body $B$ meson decays \cite{Liu:2019ymi, Li:2004ep}, so we will not discuss them in this work. For the $\pi\pi$-pair, its $S$-wave wave function can be written as \cite{Diehl:1998dk,Diehl:2000uv,Pire:2002ut,Xing:2019xti,Wang:2018xux}
\begin{eqnarray}
\Phi_{\pi\pi}= \frac{1}{\sqrt{2N_c}}\left[P\mkern-10.5mu/\phi_S(z,\xi,\omega)
+\omega\phi_S^s(z,\xi,\omega)+\omega(n\mkern-10.5mu/v\mkern-10.5mu/-1)
\phi_S^t(z,\xi,\omega)\right],
\end{eqnarray}
with $z$ being the momentum fraction of the light quark in the $\pi\pi$-pair. The parameter $\xi$ is the momentum fraction of one $\pi$ meson in the $\pi\pi$-pair. The momentum of the $\pi\pi$-pair $P$ satisfies the condition $P^2=\omega^2$, $\omega$ being the invariant mass of $\pi\pi$-pair.  $n=(1,0, \vec{0})$ and $v=(0,1,\vec{0})$ are the light-like vectors. For the explicit expressions of the light-core distributions $\phi_S^{(s,t)}$, we adopt the form \cite{Wang:2014ira,Wang:2015uea}
\begin{eqnarray}
&&\phi_S(z,\xi,\omega)=\frac{F_S(\omega)}{\sqrt{6}}9 a_sz(1-z)(2z-1),\nonumber\\
&&\phi_S^s(z,\xi,\omega)=\frac{F_S(\omega)}{2\sqrt{6}},\nonumber\\
&&\phi_S^t(z,\xi,\omega)=\frac{F_S(\omega)}{2\sqrt{6}}(1-2z),
\end{eqnarray}
with Gegenbauer moment $a_s=0.3\pm0.2$ \cite{Xing:2019xti}. $F_S(\omega)$ is the time-like form factor. In particular, for a narrow intermediate resonance, the time-like form factor $F_S(\omega)$ can be well described by the relative Breit-Wigner lineshape \cite{Back:2017zqt}. However, due to the remarkable interference between two decays  $f_0 \to \pi \pi$ and $f_0 \to K \overline{K}$, the relative Breit-Wigner lineshape cannot work well for the time-like form factor of $f_0$.  In this case, the Flatt$\acute{e}$ lineshape is proposed to describe that of $f_0$ \cite{Flatte:1976xu,Back:2017zqt}, which is given as
\begin{eqnarray}
F_S(\omega)=\frac{m_{f_0}^2}{m^2_{f_0}-\omega^2-i m_{f_0}(g_{\pi\pi}\rho_{\pi\pi}+g_{KK}\rho_{KK}F^2_{KK})},
\end{eqnarray}
with
\begin{eqnarray}
\rho_{\pi\pi}=\sqrt{1-\frac{4m^2_{\pi}}{\omega^2}},\,\,\,
\rho_{KK}=\sqrt{1-\frac{4m_K^2}{\omega^2}}.
\end{eqnarray}
The $g_{\pi\pi}$ and $g_{KK}$ are the coupling constants corresponding to $f_0\to \pi\pi$ and $f_0\to K\overline{K}$ decays, respectively, whose values are taken as $g_{\pi\pi}=(0.165\pm0.018)~\mathrm{GeV}^2$ and $g_{KK}/g_{\pi\pi}=4.21\pm0.33$ \cite{Back:2017zqt}. In addition, the factor $F_{KK}= e^{-\alpha q^2}$ is introduced to suppress the $f_0$ width above the $KK$ threshold. The parameter $\alpha$ is taken $2.0\pm1.0~\mathrm{GeV}^{-2}$, which does not affect the predictions remarkably \cite{Aaij:2014emv}. It is noted that this lineshape has been also adopted extensively in analyzing data in the LHCb experiment \cite{Aaij:2014emv}.

Although the quark model has achieved great successes, the underlying structures of the scalar mesons are not well established so far. There are many scenarios for the classification of the scalar mesons. One scenario is the naive 2-quark model, and the light scalar mesons below or near 1 GeV are identified as the lowest lying states. Another consistent picture \cite{Close:2002zu} provided by the data implies that light scalar mesons below or near 1 GeV can be described by the $q^2\bar{q}^2$, while scalars above 1 GeV will form a conventional $q\bar{q}$ nonet with with some possible glue content \cite{Cheng:2005ye, Jaffe:1976ig, Alford:2000mm}. This picture can be used to interpret the mass degeneracy of $f_0$ and $a_0(980)$, the reason why the widths of $\kappa(800)$ and $\sigma(600)$ is broader than those of $a_0(980)$ and $f_0$, and  the large couplings of $f_0$ and $a_0(980)$ to $K\overline{K}$. However, in practice it is hard for us to make quantitative predictions on $B$ decays based on the four-quark picture for light scalar mesons as it involves the unknown form factors and decay constants that are beyond the conventional quark model. Hence, we here only discuss the two-quark scenario for $f_0$. Moreover, some experimental evidences indicate the existence of the non-strange and strange quark contents in $f_0$, we therefore regard it as a mixture of $s\bar s$ and $n\bar n=(u\bar{u}+d\bar{d})/\sqrt{2}$
\begin{eqnarray}\label{mixing}
|f_0\rangle&=&|n\bar{n}\rangle\sin\theta+|s\bar{s}\rangle\cos\theta,
\end{eqnarray}
where $\theta$ is the mixing angle. Recent studies \cite{Cheng:2002ai, Anisovich:2002wy, Gokalp:2004ny} show that the mixing angle $\theta$ lies in the ranges of $25^{\circ}<\theta<40^{\circ}$ and $140^{\circ}<\theta<165^{\circ}$, and studies based on the $B$ decays favor the later range.

\begin{figure}[!htb]
\begin{center}
\includegraphics[scale=0.5]{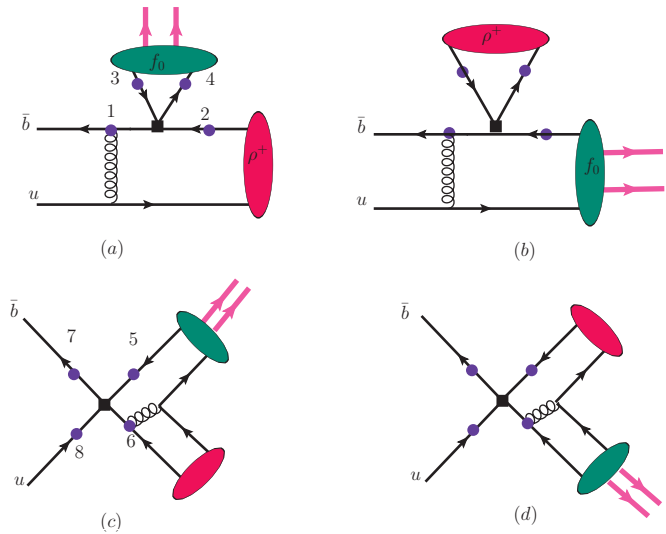}
\caption{Typical Feynman diagrams for the quasi-two-body decay $B^+\to \rho^+  \pi^+\pi^-$ in PQCD, where the black squares stand for the weak vertices, and large (purple) spots on the quark lines denote possible attachments of hard gluons. The green ellipse represent $\pi\pi$-pair and the red one is the light bachelor $ \rho^+$ meson.}\label{feynman}
\end{center}
\end{figure}

With the initial and final wave functions, we can calculate the whole amplitude of each decay mode in PQCD approach. In the leading order, the diagrams contributing to the decay $B^+\to \rho^+ \pi^+\pi^-$ are shown in the Fig.\ref{feynman}. The first two diagrams are the emission type diagrams with the first one emitting the $\pi\pi$-pair and the second one with the vector meson emitted. The last two are the annihilation type diagrams. Because the decay amplitudes are very similar to those presented in the ref.\cite{Zou:2020dpg}, for the sake of simplicity, we shall not present them in this work. The other parameters used in the numerical calculations, such as the mass of the mesons, CKM matrix elements and the life times of $B$ mesons, are taken from the Particle Data Group \cite{Zyla:2020zbs}.

\section{Numerical Results}
\begin{figure*}[!ht]
\vspace{-1.2cm}
\includegraphics[width=7cm,height=6cm]{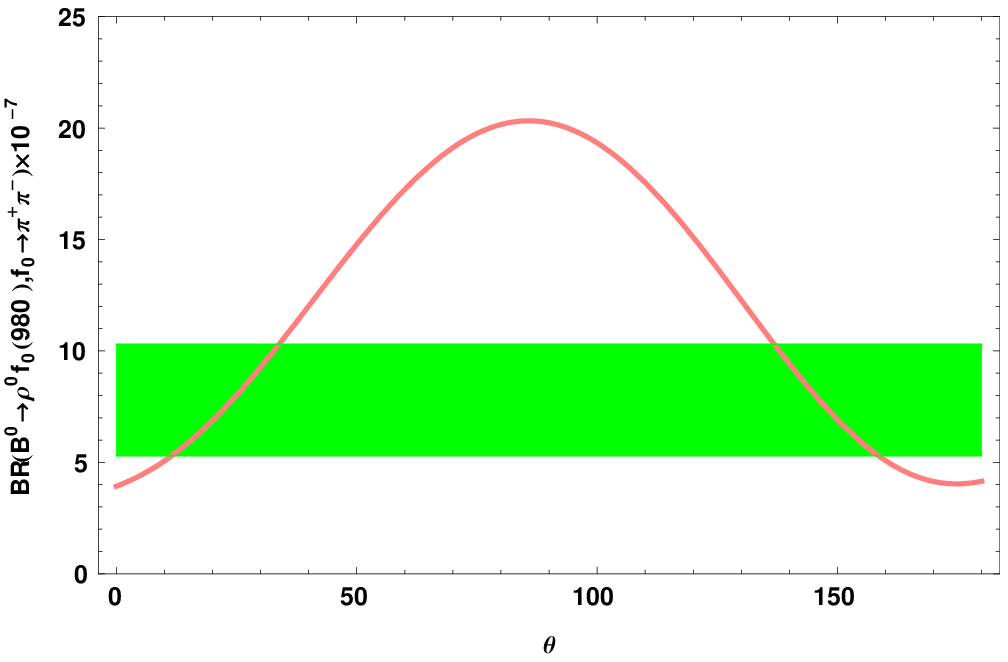}\;\;\;\;
\includegraphics[width=7cm,height=6cm]{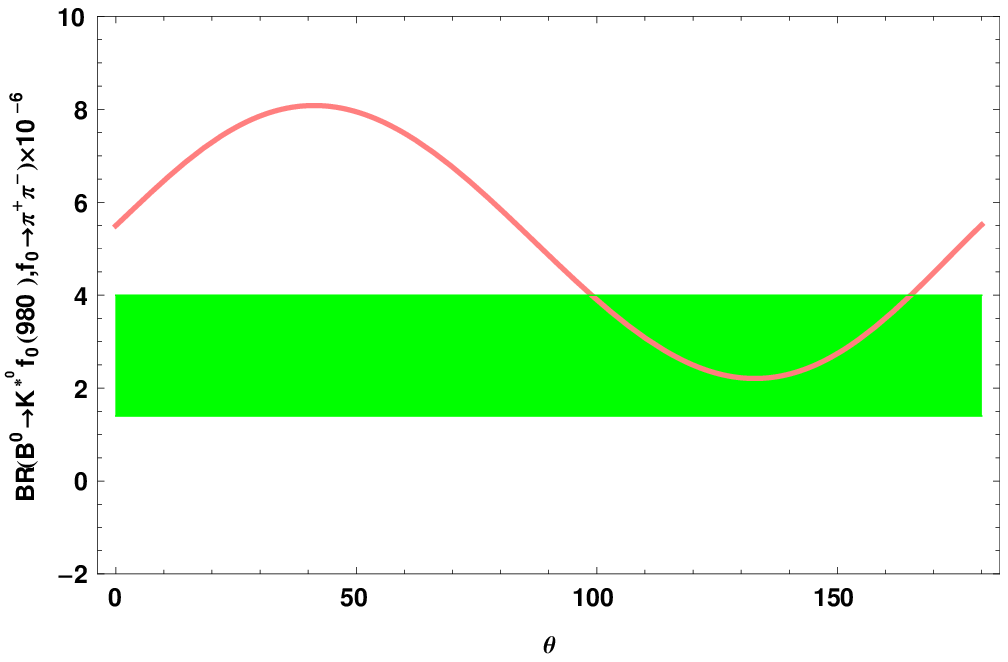}\\
\vspace{-1.5cm}
\includegraphics[width=7cm,height=6cm]{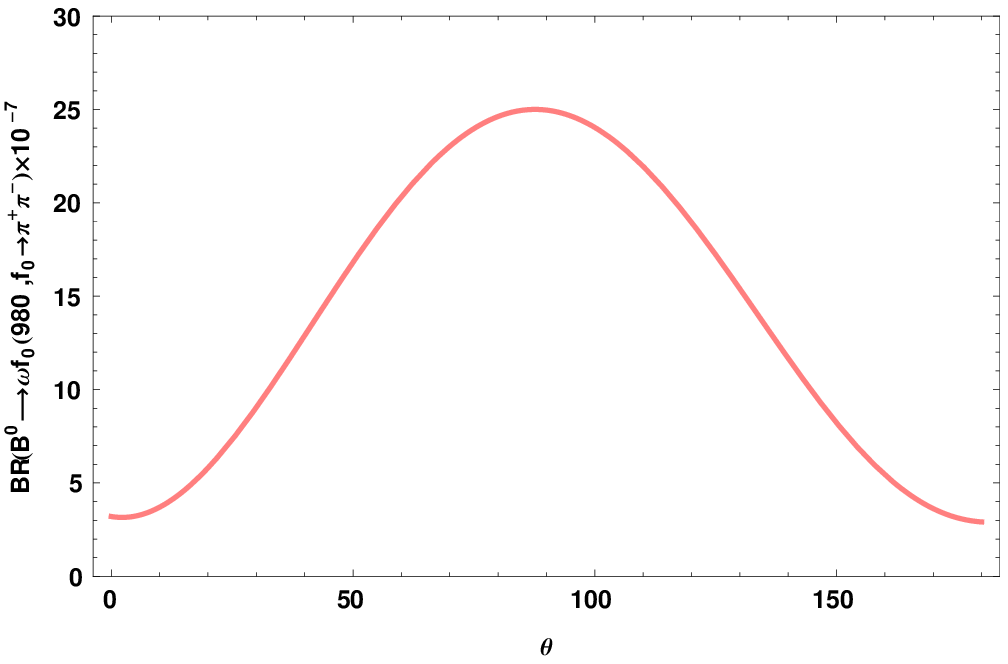}\;\;\;\;
\includegraphics[width=7cm,height=6cm]{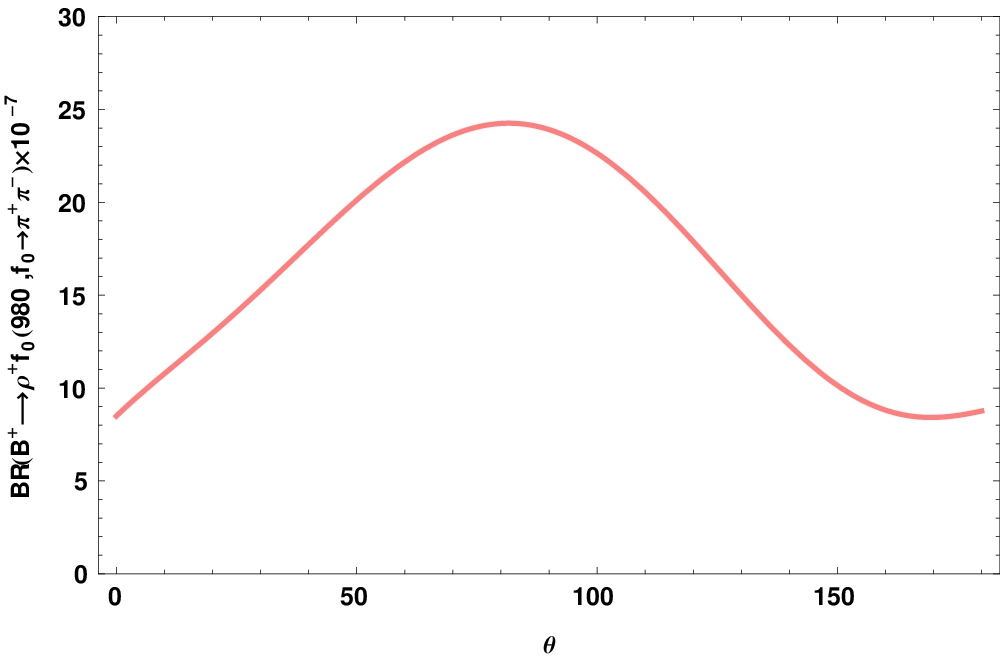}\\
\vspace{-1.5cm}
\includegraphics[width=7cm,height=6cm]{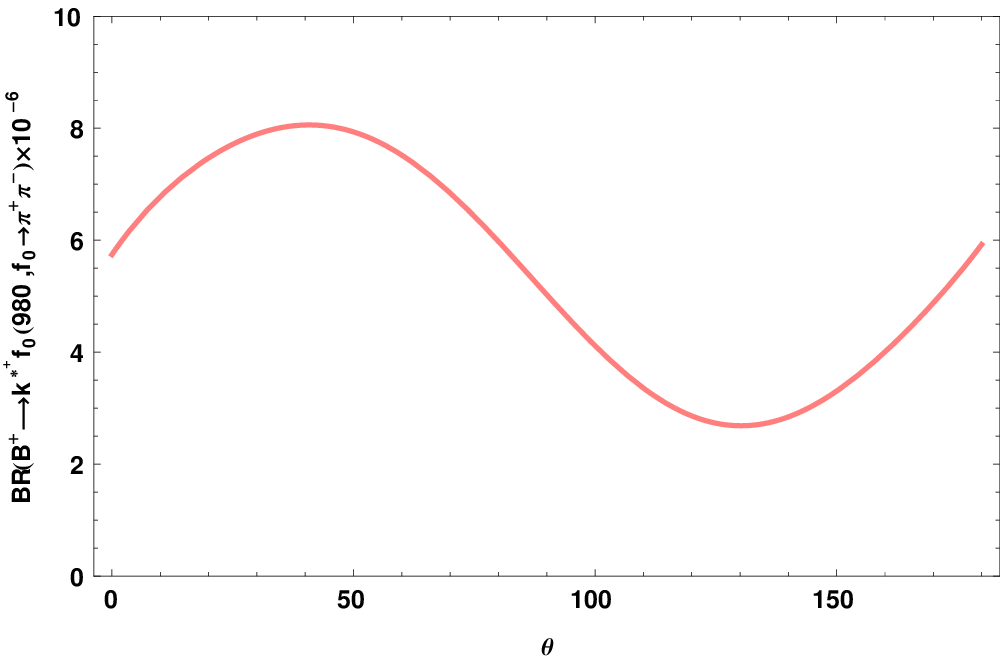}\;\;\;\;
\includegraphics[width=7cm,height=6cm]{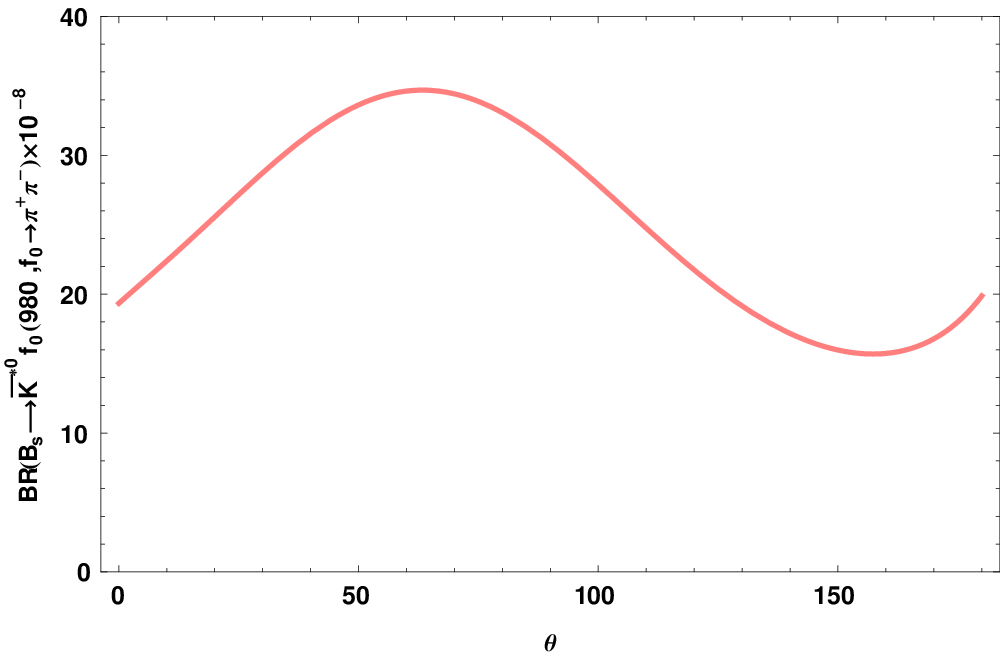}\\
\vspace{-1.5cm}
\includegraphics[width=7cm,height=6cm]{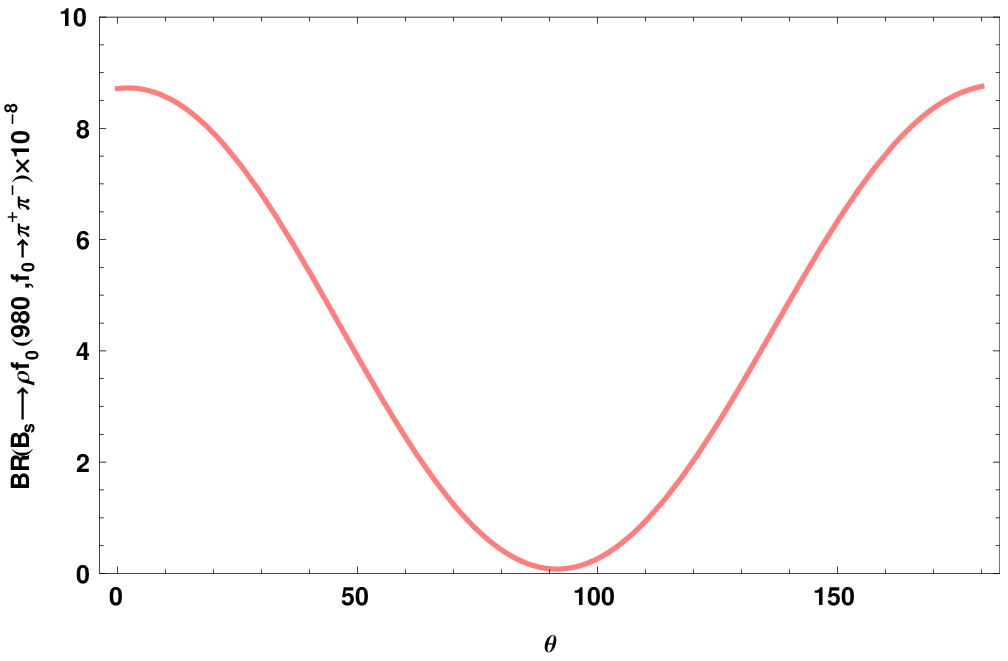}\;\;\;\;
\includegraphics[width=7cm,height=6cm]{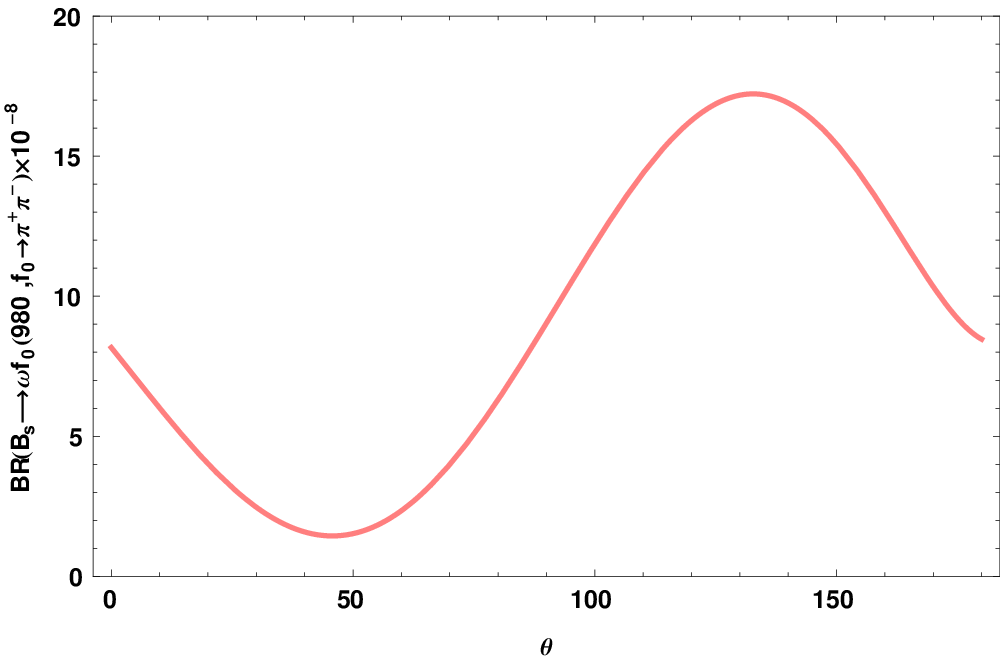}
\caption{Dependence of the $CP$-averaged  branching fractions of the quasi-two-body $B_{(s)}\to V \pi^+\pi^-$ decays on the mixing angle $\theta$ in the PQCD approach. The $\theta$ dependence of the branching fractions. The green bands are the allowed regions in the experiments.}\label{Fig:smresults}
\end{figure*}

As aforementioned, the mixing angle of $f_0$ have not yet been determined. At first, we set $\theta$ as a free parameter and plot the branching fractions of $B_{u,d,s}\to V f_0 \to V \pi^+\pi^-$ decays dependent on it in Fig.~\ref{Fig:smresults}, where the green bands are the allowed regions in the experiments. Combining the experimental data of $B^0\to \rho^0(f_0\to)\pi^+\pi^-$ and $B^0\to K^*(892)^0(f_0\to)\pi^+\pi^-$ decays, we get the range of the mixing angle $135^{\circ} \leq \theta \leq 155^{\circ}$, which is consistent with the results obtained from $\phi\to f_0\gamma$ and $f_0\to \gamma\gamma$.  In our previous work \cite{Li:2019jlp},  the decays $B\to K^*_{0,2}(1430)f_0/\sigma$ have been investigated,  and we obtained $\theta \approx145^{\circ}$ after comparing with experimental result \cite{Lees:2011dq}. If the mixing angle $\theta=145^{\circ}$ is adopted, the branching ratios of the $B^0\to \rho^0(f_0\to)\pi^+\pi^-$ and $B^0\to K^*(892)^0(f_0\to)\pi^+\pi^-$ decays are given as
\begin{eqnarray}
\mathcal{B}(B^0\to \rho^0(770)(f_0\to)\pi^+\pi^-)&=&8.25\times10^{-7},\\
\mathcal{B}(B^0\to K^*(892)^0(f_0\to)\pi^+\pi^-)&=&2.45\times10^{-6},
\end{eqnarray}
which well match the experimental measurements:
\begin{eqnarray}
\mathcal{B}(B^0\to \rho^0(770)(f_0\to)\pi^+\pi^-)&=&(7.8\pm2.5)\times10^{-7},\\
\mathcal{B}(B^0\to K^*(892)^0(f_0\to)\pi^+\pi^-)&=&(2.6_{-1.2}^{+1.4})\times10^{-6}.
\end{eqnarray}
In view of this, we present all calculated results of the $CP$-averaged branching fractions and the local direct $CP$ asymmetries of the concerned decay modes with $\theta=145^{\circ}$ in Table.~\ref{br}. For comparison, the available experimental data are also listed. One can find that adopting the appropriate wave functions of initial and final states, our predictions are in good agreement with the current experimental data, although there are only upper limits for the $B^0\to \omega(f_0\to)\pi^+\pi^-$ and $B^+\to \rho^+(f_0\to)\pi^+\pi^-$ decays.  Finally, we plot all the branching fractions dependent on the mixing angle $\theta$, which may shed light on the mixing angle by combining the ongoing experimental measurements.

\begin{table*}[!t]
\caption{The results of $CP$ averaged branching fractions (in $10^{-6}$) and the direct $CP$ asymmetries (\%) in PQCD approach.}
 \label{br}
\begin{tabular}{c c c c}
 \hline \hline
Decay Modes&Br(PQCD)&Br(EXP) \cite{Zyla:2020zbs} &$A_{CP}^{dir}$ \\
\hline
\hline
 $B^0 \to \rho^0(f_0\to)\pi^+ \pi^-$
 &$0.82^{+0.36+0.02+0.05}_{-0.34-0.16-0.10}$
 &$0.78\pm0.25$
 &$-11.4_{-3.54-8.72-0.00}^{+14.0+23.6+9.77}$\\

 $B^0 \to K^{*0}(f_0\to)\pi^+ \pi^-$
 &$2.45^{+0.66+0.56+0.00}_{-1.28-1.04-0.35}$
 &$2.6_{-1.2}^{+1.4}$
 &$-5.97_{-0.57-0.00-0.00}^{+13.7+6.72+2.82}$ \\

 $B^0 \to \omega(f_0\to)\pi^+ \pi^-$
 &$0.97^{+0.51+0.16+0.13}_{-0.39-0.19-0.10}$
 &$<1.5 $
 &$-13.7_{-3.15-14.9-0.00}^{+6.10+11.5+1.99}$ \\

 $B^+ \to \rho^+(f_0\to)\pi^+ \pi^- $
 &$1.23^{+0.50+0.25+0.00}_{-0.76-0.38-0.12}$
 &$<2.0$
 &$-55.9_{-19.5-4.34-7.33}^{+15.4+31.5+2.50}$\\

 $B^+\to K^{*+}(f_0\to)\pi^+\pi^-$
 &$3.18^{+0.94+0.76+0.00}_{-1.48-1.11-0.39}$
 &...
 &$-26.4_{-6.34-1.21-3.05}^{+9.87+4.18+3.05}$\\

 $B_s\to \rho^0(f_0\to)\pi^+\pi^-$
 &$0.06_{-0.02-0.01-0.00}^{+0.02+0.01+0.00}$
 & ...
 &$8.82^{+4.23+1.68+0.00}_{-3.51-0.72-0.02}$\\

 $B_s\to \omega (f_0\to)\pi^+\pi^-$
 &$0.17_{-0.09-0.04-0.02}^{+0.10+0.02+0.00}$
 &...
 &$14.8_{-13.1-8.17-0.53}^{+9.43+11.1+2.56}$\\

 $B_s\to \bar{K}^{*0}(f_0\to)\pi^+\pi^-$
 &$0.15^{+0.09+0.05+0.01}_{-0.04-0.02-0.00}$
 &...
 &$87.7_{-22.1-14.3-7.51}^{+0.00+0.00+0.00}$\\
 \hline
 \hline
\end{tabular}
\end{table*}

We acknowledge that there are many uncertainties in our calculations. In this work, we mainly evaluate three kinds of errors, namely the parameters in wave functions of the initial and final states, the higher order and power corrections, and the CKM matrix elements, respectively. In fact, the first kind of errors come from the uncertainties of nonperturbative parameters, such as the $B_{(s)}$ meson decay constants $f_B/f_{B_s}=0.19 \pm 0.02/0.23 \pm 0.02~\mathrm{GeV}$, the sharp parameters $\omega/\omega_s=0.4\pm0.04/0.5\pm0.05 ~\mathrm{GeV}$ in the distribution amplitudes of $B$ mesons, the Gegenbauer moments in the LCDAs of vector mesons, and the Gagenbauer moment $a_S$ in the $S$-wave LCDAs of the $\pi\pi$-pair, et.al. We can find from the table that this kind of uncertainties are dominant. Fortunately, this kind of uncertainties could be reduced with the developments of the experiments or other nonperturbative theoretical approaches in future. The second kind of errors arise from the unknown higher order of $\alpha_s$ and higher power corrections, which are reflected by varying the $\Lambda_{QCD}=0.25\pm0.05~\mathrm{GeV} $ and the factorization scale $t$ from $0.8t$ to $1.2t$, respectively. The last ones are caused by the uncertainties of the CKM matrix elements.

If the narrow-width approximation (NWA) holds in these decays, the branching fraction of the quasi-two-body $B$ meson decay can be decomposed as
\begin{eqnarray}
\mathcal{B}(B\to M_1(R\to)M_2M_3)\simeq\mathcal{B}(B\to M_1R)\times \mathcal{B}(R\to M_2M_3),
\end{eqnarray}
with $R$ represents a resonance. If two decays have a same resonance, we then define a ratio as
\begin{eqnarray}
R_{V_1/V_2}=\frac{\mathcal{B}(B\to V_1R)}{\mathcal{B}(B\to V_2R)}=\frac{\mathcal{B}(B\to V_1R)\times \mathcal{B}(R\to\pi^+\pi^-)}{\mathcal{B}(B\to V_2R)\times \mathcal{B}(R\to\pi^+\pi^-)}
\simeq\frac{\mathcal{B}(B\to V_1(R\to)\pi^+\pi^-)}{\mathcal{B}(B\to V_2(R\to)\pi^+\pi^-)}.
\end{eqnarray}
Based on the predictions in Table.~\ref{br}, the ratio between the $B^0\to \rho^0 f_0$ and $B^0\to \omega f_0$ is given as
 \begin{eqnarray}
 R_{\rho^0/\omega}=\frac{\mathcal{B}(B\to \rho^0 f_0)}{\mathcal{B}(B\to \omega f_0)}\sim 1,
 \end{eqnarray}
which is agreement with the results of QCDF \cite{Cheng:2013fba}. Within the isospin relation
\begin{eqnarray}
r=\frac{\mathcal{B}(f_0\to\pi^+\pi^-)}{\mathcal{B}(f_0\to\pi^0\pi^0)}=2,
\end{eqnarray}
we obtain the relation
 \begin{eqnarray}
 \mathcal{B}(B\to V(f_0\to)\pi^0\pi^0)=\frac{1}{2}\mathcal{B}(B\to V(f_0\to)\pi^+\pi^-),
 \end{eqnarray}
which can be used to predict the branching fractions of the corresponding quasi-two-body $B\to V(f_0\to)\pi^0\pi^0$ decays.

Now we turn to discuss the local direct $CP$ asymmetries of these decays. In the quark level,  $B\to K^{*}(f_0\to)\pi^+\pi^-$, $B_s\to \rho^0/\omega(f_0\to)\pi^+\pi^-$, and $B_s\to \phi(f _0(980)\to)\pi^+\pi^-$ are induced by $\bar b\to \bar s q\bar q$ transition, while $B\to (\rho, \omega)(f_0\to)\pi^+\pi^-$ and $B_s\to K^{*}(f _0(980)\to)\pi^+\pi^-$ are controlled by $ \bar b\to \bar d q\bar q$ transition. From Table.~\ref{br}, it is found that the local $CP$ asymmetries of decays $B^{0}\to K^{*0}(f_0\to)\pi^+\pi^-$ and $B_s\to \rho^0/\omega(f_0\to)\pi^+\pi^-$ are very small, and the reason is that the tree diagrams contributions are both color and CKM elements suppressed. However, for the decay $B^+\to K^{*+}(f_0\to)\pi^+\pi^-$, because the spectator $u$ quark  enters into not only $K^{*+}$ meson but also $\pi\pi$-pair, the contributions from tree and penguin operators are comparable, leading to a large $CP$ asymmetry. For the decays $B^0\to \rho^0/\omega(f_0\to)\pi^+\pi^-$, although the contributions from tree operators are color suppressed, the destructive interference between the diagrams with vector meson emitted and ones with $\pi\pi$-pair emitted decreases the effects of penguin operators remarkably, therefore the $CP$ asymmetries are as small as about $10\%$. For $B^+\to \rho^+(f_0\to)\pi^+\pi^-$ decay, its amplitude is more complicated. In eq.~(\ref{mixing}), if the mixing angle $\theta$ of $f_0 (980) $ is an obtuse angle, the sign of $n\bar n$ is negative. The spectator $u$-quark of $B^+\to \rho^+(f_0\to)\pi^+\pi^-$ can enter into both $f_0$ and $\pi\pi$-pair, so the negative sign leads to the cancellation between two tree operators contributions. With the sizable contributions of penguin operators, the $CP$ asymmetry of this decay is as large as $-55\%$. The decay of $B_s\to \bar{K}^{*0}(f_0\to)\pi^+\pi^-$ is very similar to $B^+\to \rho^+(f_0\to)\pi^+\pi^-$, but the spectator is a strange quark.  When the spectator enters into the kaon, both the tree and penguin operators contribute, and the tree operators are color suppressed. However, when it enters into the $\pi\pi$-pair, only penguin operators play roles. Due to large interference between two kinds of above contributions, the large $CP$ asymmetry in $B_s\to \bar{K}^{*0}(f_0\to)\pi^+\pi^-$ is reasonable. On the experimental side, these $CP$ asymmetries have not been measured, and we hope these predictions can be tested in future.
\section{Summary}
In this work, we investigated the quasi-two-body $B/B_s\to V(f_0(980)\to)\pi^+\pi^-$ decays in PQCD approach, assuming that $f_0(980)$ is a mixture of $ n\bar n=(u\bar u+d\bar d)/{\sqrt 2} $ and $s \bar s$ with the mixing angle $\theta$. Within the $S$-wave two-pion wave function, both the branching fractions and the located $CP$ asymmetries have been calculated.  When the mixing angle $\theta$ is around $145^{\circ}$, the obtained branching fractions of the $B^0\to \rho^0(f_0(980)\to)\pi^+\pi^-$ and $B^0\to K^{*0}(f_0(980)\to)\pi^+\pi^-$ are in good agreement with the experimental data, and other results could be tested in the future experiments. In addition, the branching fractions of $B/B_s\to V(f_0\to)\pi^0\pi^0$ could be predicted based on the isospin symmetry, which can be measured in the LHCb and Belle II experiments. We note that the calculated $CP$ asymmetries of the $B^+\to \rho^+(f_0\to)\pi^+\pi^-$ and $B_s\to \bar{K}^{*0}(f_0\to)\pi^+\pi^-$ decays are very large, which can be tested in the ongoing experiments. We acknowledge that there are many uncertainties in the calculation, and the dominant one is the two-meson wave function. Therefore, more precise multi-meson wave functions from non-perturbative approach are needed.

\section*{Acknowledgment}
This work is supported in part by the National Science Foundation of China under the Grant Nos. 11705159, 11975195, 11875033, and 11765012, and the Natural Science Foundation of Shandong province under the Grant No.ZR2019JQ04. This work is also supported by the Project of Shandong Province Higher Educational Science and Technology Program under Grants No. 2019KJJ007.
\bibliographystyle{bibstyle}
\bibliography{mybibfile}

\providecommand{\href}[2]{#2}\begingroup\raggedright\begin{thebibliography}{10}

\bibitem{Lees:2012kxa}
{\bf BaBar} Collaboration, J.~P. Lees et~al., {\it {Study of CP violation in
  Dalitz-plot analyses of $B^0 \to K^+K^-K^0_S$, $B^+\to K^+K^-K^+$, and
  $B^+\to K^0_S K^0_SK^+$}},  {\em Phys. Rev.} {\bf D85} (2012) 112010,
  [\href{https://arxiv.org/abs/1201.5897}{{\tt arXiv:1201.5897}}].

\bibitem{Nakahama:2010nj}
{\bf Belle} Collaboration, Y.~Nakahama et~al., {\it {Measurement of CP
  violating asymmetries in $B^0 \to K^+K^- K^0_S$ decays with a time-dependent
  Dalitz approach}},  {\em Phys. Rev.} {\bf D82} (2010) 073011,
  [\href{https://arxiv.org/abs/1007.3848}{{\tt arXiv:1007.3848}}].

\bibitem{Eckhart:2002qr}
{\bf CLEO} Collaboration, E.~Eckhart et~al., {\it {Observation of $B\to K^0_S
  \pi^+\pi^-$ and evidence for $B\to K^{*\pm}\pi^\mp$}},  {\em Phys. Rev.
  Lett.} {\bf 89} (2002) 251801,
  [\href{https://arxiv.org/abs/hep-ex/0206024}{{\tt hep-ex/0206024}}].

\bibitem{Aaij:2018rol}
{\bf LHCb} Collaboration, R.~Aaij et~al., {\it {Observation of the decay $B_s^0
  \to \overline{D}^0 K^+ K^-$}},  {\em Phys. Rev.} {\bf D98} (2018), no.~7
  072006, [\href{https://arxiv.org/abs/1807.01891}{{\tt arXiv:1807.01891}}].

\bibitem{Aaij:2016qnm}
{\bf LHCb} Collaboration, R.~Aaij et~al., {\it {Observation of the decay $B^0_s
  \to \phi\pi^+\pi^-$ and evidence for $B^0 \to \phi\pi^+\pi^-$}},  {\em Phys.
  Rev.} {\bf D95} (2017), no.~1 012006,
  [\href{https://arxiv.org/abs/1610.05187}{{\tt arXiv:1610.05187}}].

\bibitem{Aaij:2019nmr}
{\bf LHCb} Collaboration, R.~Aaij et~al., {\it {Amplitude analysis of
  $B^{0}_{s} \to K^{0}_{\textrm{S}} K^{\pm}\pi^{\mp}$ decays}},  {\em JHEP}
  {\bf 06} (2019) 114, [\href{https://arxiv.org/abs/1902.07955}{{\tt
  arXiv:1902.07955}}].

\bibitem{Aaij:2017zgz}
{\bf LHCb} Collaboration, R.~Aaij et~al., {\it {Resonances and $CP$ violation
  in $B_s^0$ and $\overline{B}_s^0 \to J/\psi K^+K^-$ decays in the mass region
  above the $\phi(1020)$}},  {\em JHEP} {\bf 08} (2017) 037,
  [\href{https://arxiv.org/abs/1704.08217}{{\tt arXiv:1704.08217}}].

\bibitem{Aaij:2019jaq}
{\bf LHCb} Collaboration, R.~Aaij et~al., {\it {Amplitude analysis of the $B^+
  \rightarrow \pi^+\pi^+\pi^-$ decay}},  {\em Phys. Rev.} {\bf D101} (2020),
  no.~1 012006, [\href{https://arxiv.org/abs/1909.05212}{{\tt
  arXiv:1909.05212}}].

\bibitem{Aaij:2020ypa}
{\bf LHCb} Collaboration, R.~Aaij et~al., {\it {Amplitude analysis of the
  $B^+\to D^+D^-K^+$ decay}},  {\em Phys. Rev. D} {\bf 102} (2020) 112003,
  [\href{https://arxiv.org/abs/2009.00026}{{\tt arXiv:2009.00026}}].

\bibitem{Aaij:2020dsq}
{\bf LHCb} Collaboration, R.~Aaij et~al., {\it {Measurement of the relative
  branching fractions of $B^+ \to h^+h^{\prime +}h^{\prime -}$ decays}},  {\em
  Phys. Rev. D} {\bf 102} (2020) 112010,
  [\href{https://arxiv.org/abs/2010.11802}{{\tt arXiv:2010.11802}}].

\bibitem{He:2014xha}
X.-G. He, G.-N. Li, and D.~Xu, {\it {SU(3) and isospin breaking effects on $B
  \to PPP$ amplitudes}},  {\em Phys. Rev.} {\bf D91} (2015), no.~1 014029,
  [\href{https://arxiv.org/abs/1410.0476}{{\tt arXiv:1410.0476}}].

\bibitem{Krankl:2015fha}
S.~Krankl, T.~Mannel, and J.~Virto, {\it {Three-body non-leptonic B decays and
  QCD factorization}},  {\em Nucl. Phys.} {\bf B899} (2015) 247--264,
  [\href{https://arxiv.org/abs/1505.04111}{{\tt arXiv:1505.04111}}].

\bibitem{Virto:2016fbw}
J.~Virto, {\it {Charmless Non-Leptonic Multi-Body B decays}},  {\em PoS} {\bf
  FPCP2016} (2017) 007, [\href{https://arxiv.org/abs/1609.07430}{{\tt
  arXiv:1609.07430}}].

\bibitem{Cheng:2007si}
H.-Y. Cheng, C.-K. Chua, and A.~Soni, {\it {Charmless three-body decays of B
  mesons}},  {\em Phys. Rev.} {\bf D76} (2007) 094006,
  [\href{https://arxiv.org/abs/0704.1049}{{\tt arXiv:0704.1049}}].

\bibitem{Cheng:2016shb}
H.-Y. Cheng, C.-K. Chua, and Z.-Q. Zhang, {\it {Direct CP Violation in
  Charmless Three-body Decays of $B$ Mesons}},  {\em Phys. Rev.} {\bf D94}
  (2016), no.~9 094015, [\href{https://arxiv.org/abs/1607.08313}{{\tt
  arXiv:1607.08313}}].

\bibitem{Li:2014oca}
Y.~Li, {\it {Comprehensive study of $\overline B^0\to K^0(\overline K^0)
  K^\mp\pi^\pm$ decays in the factorization approach}},  {\em Phys. Rev.} {\bf
  D89} (2014), no.~9 094007, [\href{https://arxiv.org/abs/1402.6052}{{\tt
  arXiv:1402.6052}}].

\bibitem{Wang:2016rlo}
W.-F. Wang and H.-n. Li, {\it {Quasi-two-body decays $B\to K\rho\to K\pi\pi$ in
  perturbative QCD approach}},  {\em Phys. Lett.} {\bf B763} (2016) 29--39,
  [\href{https://arxiv.org/abs/1609.04614}{{\tt arXiv:1609.04614}}].

\bibitem{Li:2016tpn}
Y.~Li, A.-J. Ma, W.-F. Wang, and Z.-J. Xiao, {\it {Quasi-two-body decays
  $B_{(s)}\to P\rho\to P\pi\pi$ in perturbative QCD approach}},  {\em Phys.
  Rev.} {\bf D95} (2017), no.~5 056008,
  [\href{https://arxiv.org/abs/1612.05934}{{\tt arXiv:1612.05934}}].

\bibitem{Zou:2020atb}
Z.-T. Zou, Y.~Li, Q.-X. Li, and X.~Liu, {\it {Resonant contributions to
  three-body $B\rightarrow KKK$ decays in perturbative QCD approach}},  {\em
  Eur. Phys. J. C} {\bf 80} (2020), no.~5 394,
  [\href{https://arxiv.org/abs/2003.03754}{{\tt arXiv:2003.03754}}].

\bibitem{Zou:2020fax}
Z.-T. Zou, Y.~Li, and X.~Liu, {\it {Branching fractions and CP asymmetries of
  the quasi-two-body decays in $B_{s} \rightarrow K^0({\overline{K}}^0)K^\pm
  \pi ^\mp $ within PQCD approach}},  {\em Eur. Phys. J. C} {\bf 80} (2020),
  no.~6 517, [\href{https://arxiv.org/abs/2005.02097}{{\tt arXiv:2005.02097}}].

\bibitem{Zou:2021lex}
Z.-T. Zou, Y.~Li, and H.-n. Li, {\it {Is $f_X(1500)$ observed in the $B\to
  \pi(K)KK$ decays $\rho^0(1450)$?}},  {\em Phys. Rev. D} {\bf 103} (2021),
  no.~1 013005, [\href{https://arxiv.org/abs/2007.13141}{{\tt
  arXiv:2007.13141}}].

\bibitem{Zou:2020dpg}
Z.-T. Zou, L.~Yang, Y.~Li, and X.~Liu, {\it {Study of quasi-two-body
  $B_{(s)}\rightarrow \phi (f_0(980)/f_2(1270)\rightarrow )\pi \pi $ decays in
  perturbative QCD approach}},  {\em Eur. Phys. J. C} {\bf 81} (2021), no.~1
  91, [\href{https://arxiv.org/abs/2011.07676}{{\tt arXiv:2011.07676}}].

\bibitem{Wang:2015ula}
C.~Wang, Z.-H. Zhang, Z.-Y. Wang, and X.-H. Guo, {\it {Localized direct CP
  violation in $B^\pm \rightarrow \rho ^0 (\omega )\pi ^\pm \rightarrow \pi ^+
  \pi ^-\pi ^\pm $}},  {\em Eur. Phys. J.} {\bf C75} (2015), no.~11 536,
  [\href{https://arxiv.org/abs/1506.00324}{{\tt arXiv:1506.00324}}].

\bibitem{Zyla:2020zbs}
{\bf Particle Data Group} Collaboration, P.~Zyla et~al., {\it {Review of
  Particle Physics}},  {\em PTEP} {\bf 2020} (2020), no.~8 083C01.

\bibitem{Li:2001ay}
H.-n. Li, {\it {Threshold resummation for exclusive $B$ meson decays}},  {\em
  Phys. Rev. D} {\bf 66} (2002) 094010,
  [\href{https://arxiv.org/abs/hep-ph/0102013}{{\tt hep-ph/0102013}}].

\bibitem{Lu:2000hj}
C.-D. Lu and M.-Z. Yang, {\it {$B \to \pi \rho, \pi \omega$ decays in
  perturbative QCD approach}},  {\em Eur. Phys. J.} {\bf C23} (2002) 275--287,
  [\href{https://arxiv.org/abs/hep-ph/0011238}{{\tt hep-ph/0011238}}].

\bibitem{Liu:2019ymi}
X.~Liu, Z.-T. Zou, Y.~Li, and Z.-J. Xiao, {\it {Phenomenological studies on the
  $B_{d,s}^0 \to J/\psi f_0(500) [f_0(980)]$ decays}},  {\em Phys.\ Rev.\ D}
  {\bf 100} (2019), no.~1 013006, [\href{https://arxiv.org/abs/1906.02489}{{\tt
  arXiv:1906.02489}}].

\bibitem{Li:2004ep}
Y.~Li, C.-D. Lu, Z.-J. Xiao, and X.-Q. Yu, {\it {Branching ratio and CP
  asymmetry of $B_s \to\pi^+\pi^-$ decays in the perturbative QCD approach}},
  {\em Phys. Rev.} {\bf D70} (2004) 034009,
  [\href{https://arxiv.org/abs/hep-ph/0404028}{{\tt hep-ph/0404028}}].

\bibitem{Diehl:1998dk}
M.~Diehl, T.~Gousset, B.~Pire, and O.~Teryaev, {\it {Probing partonic structure
  in $\gamma^* \gamma \to \pi \pi$ near threshold}},  {\em Phys. Rev. Lett.}
  {\bf 81} (1998) 1782--1785, [\href{https://arxiv.org/abs/hep-ph/9805380}{{\tt
  hep-ph/9805380}}].

\bibitem{Diehl:2000uv}
M.~Diehl, T.~Gousset, and B.~Pire, {\it {Exclusive production of pion pairs in
  $ \gamma^* \gamma$ collisions at large $Q^2$}},  {\em Phys. Rev. D} {\bf 62}
  (2000) 073014, [\href{https://arxiv.org/abs/hep-ph/0003233}{{\tt
  hep-ph/0003233}}].

\bibitem{Pire:2002ut}
B.~Pire and L.~Szymanowski, {\it {Impact representation of generalized
  distribution amplitudes}},  {\em Phys. Lett. B} {\bf 556} (2003) 129--134,
  [\href{https://arxiv.org/abs/hep-ph/0212296}{{\tt hep-ph/0212296}}].

\bibitem{Xing:2019xti}
Y.~Xing and Z.-P. Xing, {\it {$S$-wave contributions in $\bar B_s^0\to
  (D^0,\bar D^0)\pi^+\pi^- $ within perturbative QCD approach}},  {\em Chin.
  Phys.} {\bf C43} (2019), no.~7 073103,
  [\href{https://arxiv.org/abs/1903.04255}{{\tt arXiv:1903.04255}}].

\bibitem{Wang:2018xux}
N.~Wang, Q.~Chang, Y.~Yang, and J.~Sun, {\it {Study of the $B_{s}$ ${\to}$
  ${\phi}f_{0}(980)$ ${\to}$ ${\phi}\,{\pi}^{+}{\pi}^{-}$ decay with
  perturbative QCD approach}},  {\em J. Phys. G} {\bf 46} (2019), no.~9 095001,
  [\href{https://arxiv.org/abs/1803.02656}{{\tt arXiv:1803.02656}}].

\bibitem{Wang:2014ira}
W.-F. Wang, H.-C. Hu, H.-n. Li, and C.-D. Lü, {\it {Direct CP asymmetries of
  three-body $B$ decays in perturbative QCD}},  {\em Phys. Rev.} {\bf D89}
  (2014), no.~7 074031, [\href{https://arxiv.org/abs/1402.5280}{{\tt
  arXiv:1402.5280}}].

\bibitem{Wang:2015uea}
W.-F. Wang, H.-n. Li, W.~Wang, and C.-D. Lü, {\it {$S$-wave resonance
  contributions to the $B^0_{(s)}\to J/\psi\pi^+\pi^-$ and
  $B_s\to\pi^+\pi^-\mu^+\mu^-$ decays}},  {\em Phys. Rev.} {\bf D91} (2015),
  no.~9 094024, [\href{https://arxiv.org/abs/1502.05483}{{\tt
  arXiv:1502.05483}}].

\bibitem{Back:2017zqt}
J.~Back et~al., {\it {LAURA$^{++}$: A Dalitz plot fitter}},  {\em Comput. Phys.
  Commun.} {\bf 231} (2018) 198--242,
  [\href{https://arxiv.org/abs/1711.09854}{{\tt arXiv:1711.09854}}].

\bibitem{Flatte:1976xu}
S.~M. Flatte, {\it {Coupled - Channel Analysis of the $\pi$ $\eta$ and $K\bar
  K$ Systems Near $K\bar K$ Threshold}},  {\em Phys. Lett.} {\bf 63B} (1976)
  224--227.

\bibitem{Aaij:2014emv}
{\bf LHCb} Collaboration, R.~Aaij et~al., {\it {Measurement of resonant and CP
  components in $\bar{B}_s^0\to J/\psi\pi^+\pi^-$ decays}},  {\em Phys. Rev.}
  {\bf D89} (2014), no.~9 092006, [\href{https://arxiv.org/abs/1402.6248}{{\tt
  arXiv:1402.6248}}].

\bibitem{Close:2002zu}
F.~E. Close and N.~A. Tornqvist, {\it {Scalar mesons above and below 1-GeV}},
  {\em J. Phys. G} {\bf 28} (2002) R249--R267,
  [\href{https://arxiv.org/abs/hep-ph/0204205}{{\tt hep-ph/0204205}}].

\bibitem{Cheng:2005ye}
H.-Y. Cheng and K.-C. Yang, {\it {$B \to f_0(980)K$ decays and subleading
  corrections}},  {\em Phys. Rev. D} {\bf 71} (2005) 054020,
  [\href{https://arxiv.org/abs/hep-ph/0501253}{{\tt hep-ph/0501253}}].

\bibitem{Jaffe:1976ig}
R.~L. Jaffe, {\it {Multi-Quark Hadrons. 1. The Phenomenology of (2 Quark 2
  anti-Quark) Mesons}},  {\em Phys. Rev. D} {\bf 15} (1977) 267.

\bibitem{Alford:2000mm}
M.~G. Alford and R.~L. Jaffe, {\it {Insight into the scalar mesons from a
  lattice calculation}},  {\em Nucl. Phys. B} {\bf 578} (2000) 367--382,
  [\href{https://arxiv.org/abs/hep-lat/0001023}{{\tt hep-lat/0001023}}].

\bibitem{Cheng:2002ai}
H.-Y. Cheng, {\it {Hadronic D decays involving scalar mesons}},  {\em Phys.
  Rev.} {\bf D67} (2003) 034024,
  [\href{https://arxiv.org/abs/hep-ph/0212117}{{\tt hep-ph/0212117}}].

\bibitem{Anisovich:2002wy}
{\bf SIGMA-AYAKS} Collaboration, A.~V. Anisovich, V.~V. Anisovich, V.~N.
  Markov, and N.~A. Nikonov, {\it {Radiative decays and quark content of
  $f_0(980)$ and $\phi(1020)$}},  {\em Phys. Atom. Nucl.} {\bf 65} (2002)
  497--512.

\bibitem{Gokalp:2004ny}
A.~Gokalp, Y.~Sarac, and O.~Yilmaz, {\it {An Analysis of $f_0-\sigma$ mixing in
  light cone QCD sum rules}},  {\em Phys. Lett. B} {\bf 609} (2005) 291--297,
  [\href{https://arxiv.org/abs/hep-ph/0410380}{{\tt hep-ph/0410380}}].

\bibitem{Li:2019jlp}
Q.-X. Li, L.~Yang, Z.-T. Zou, Y.~Li, and X.~Liu, {\it {Calculation of the
  $B\rightarrow K_{0,2}^*(1430)f_0(980)/\sigma $ decays in the perturbative QCD
  approach}},  {\em Eur. Phys. J.} {\bf C79} (2019), no.~11 960,
  [\href{https://arxiv.org/abs/1910.09209}{{\tt arXiv:1910.09209}}].

\bibitem{Lees:2011dq}
{\bf BaBar} Collaboration, J.~P. Lees et~al., {\it {$B^0$ meson decays to
  $\rho^0 K^{*0}$, $f_0 K^{*0}$, and $\rho^-K^{*+}$, including higher $K^*$
  resonances}},  {\em Phys. Rev. D} {\bf 85} (2012) 072005,
  [\href{https://arxiv.org/abs/1112.3896}{{\tt arXiv:1112.3896}}].

\bibitem{Cheng:2013fba}
H.-Y. Cheng, C.-K. Chua, K.-C. Yang, and Z.-Q. Zhang, {\it {Revisiting
  charmless hadronic B decays to scalar mesons}},  {\em Phys. Rev. D} {\bf 87}
  (2013), no.~11 114001, [\href{https://arxiv.org/abs/1303.4403}{{\tt
  arXiv:1303.4403}}].

\end{thebibliography}\endgroup
\end{document}